\newif\ifdraft
\newif\iffull
\newif\ifcomment
\newif\iflatexdiff
\newif\ifbibtex
\newif\ifpreprint
\preprinttrue
\fulltrue
\def\dvers{v1.2}
\ifpreprint
\documentclass[ALICE,manyauthors]{cernphprep}
\usepackage[comma,square,numbers,sort&compress]{natbib}
\else
\documentclass[10pt,aps,prl,twocolumn,superscriptaddress,altaffilletter,nobibnotes,nofootinbib]{revtex4-1}
\fi
\usepackage{hyperref}
\usepackage{graphicx}  
\usepackage{dcolumn}   
\usepackage{bm}        
\usepackage{amssymb}   
\usepackage{amsfonts}
\usepackage{graphics}
\usepackage{epsfig}
\usepackage[usenames]{color}
\usepackage{multirow}
\iflatexdiff
\RequirePackage[normalem]{ulem}
\RequirePackage{color}\definecolor{RED}{rgb}{1,0,0}\definecolor{BLUE}{rgb}{0,0,1}

\fi
\ifpreprint
\else

\fi

\newcommand{\ZDC}          {\rm{ZDC}}
\newcommand{\ZNA}          {\rm{ZNA}}
\newcommand{\ZNC}          {\rm{ZNC}}
\newcommand{\ZDCs}         {\rm{ZDCs}}
\newcommand{\SPD}          {\rm{SPD}}

\newcommand{\VZERO}        {\rm{VZERO}}
\newcommand{\VZEROA}       {\rm{VZERO-A}}
\newcommand{\VZEROC}       {\rm{VZERO-C}}

\newcommand{\pp}           {pp}
\newcommand{\ppbar}        {\mbox{$\mathrm {p\overline{p}}$}}
\newcommand{\PbPb}         {\mbox{Pb--Pb}}
\newcommand{\pPb}          {\mbox{p--Pb}}

\newcommand{\dAu}          {\mbox{d--Au}}
\newcommand{\pAu}          {\mbox{p--Au}}

\newcommand{\etal}         {\ensuremath{\eta_{\rm lab}}}
\newcommand{\dNdetal}      {\mathrm{d}N_\mathrm{ch}/\mathrm{d}\etal}
\newcommand{\dNdetatrl}    {\mathrm{d}N_\mathrm{tracklets}/\mathrm{d}\etal}

\newcommand{\etac}         {\ensuremath{\eta_{\rm cms}}}
\newcommand{\dNdetac}      {\mathrm{d}N_\mathrm{ch}/\mathrm{d}\etac}

\newcommand{\lum}          {\mbox{${\rm cm}^{-2} {\rm s}^{-1}$}}

\newcommand{\pt}           {\ensuremath{p_{\rm T}}}

\newcommand{\snn}          {\ensuremath{\sqrt{s_{\rm NN}}}}
\newcommand{\snnbf}        {\ensuremath{\mathbf{{\sqrt{s_{\mathbf NN}}}}}}

\newcommand{\avNpart}      {\ensuremath{\langle N_\mathrm{part} \rangle}}

\newcommand{\abs}[1]       {\ensuremath{\left|#1\right|}}

\newcommand{\stat}         {({\it stat.})}
\newcommand{\syst}         {({\it syst.})}
\newcommand{\Fig}[1]       {Fig.~\ref{#1}}

\newcommand{\Ref}[1]       {Ref.~\cite{#1}}

\newcommand{\red}[1]       {\textcolor{red}{#1}}

\newcommand{\warn}[1]      {{\small\textbf{(!\footnote{\textbf{(!)}~#1})}}\marginpar{\textbf{---}}}

\newcommand{\comment}[1]   {}

\newcommand{\final}[1]     {{\textcolor{blue}{#1}}}

\iflatexdiff

\newcommand{\old}[1]       {\red{\sout{#1}}}

\else

\newcommand{\old}[1]       {\relax}

\fi
\newcommand{\resetal}      {17.35}
\newcommand{\resetale}     {0.67}
\newcommand{\resetac}      {16.81}
\newcommand{\resetace}     {0.71}
\newcommand{\resetacp}     {2.14}
\newcommand{\resetacpe}    {0.17}
\newcommand{\facpp}        {16}

\newcommand{\facdAu}       {84}

\graphicspath{{./figs/}}
\ifdraft
\usepackage{lineno}
\linenumbers
\modulolinenumbers[5]
\usepackage{fancyhdr}
\renewcommand{\final}[1]{#1}
\else
\renewcommand{\warn}[1]{}
\renewcommand{\final}[1]{#1}
\fi
\begin{document}
\newlength{\figlen}
\setlength{\figlen}{\linewidth}
\ifpreprint
\setlength{\figlen}{0.75\textwidth}
\begin{titlepage}
\PHnumber{2012-307}                   
\PHdate{08 Oct 2012}                  
\title{Pseudorapidity density of charged particles \\in \pPb\ collisions at \snnbf\ = 5.02 TeV}
\ShortTitle{Pseudorapidity density in \pPb\ collisions}
\Collaboration{ALICE Collaboration%
         \thanks{See Appendix~\ref{app:collab} for the list of collaboration members}}
\ShortAuthor{ALICE Collaboration} 
\else
\title{Pseudorapidity density of charged particles in \pPb\ collisions at \snnbf\ = 5.02 TeV}
\iffull
\input{authors-prl.tex}
\else
\collaboration{ALICE Collaboration}
\fi
\vspace{0.3cm}
\ifdraft
\date{\today, \color{red}DRAFT \dvers\ \$Revision: 636 $\color{white}:$\$\color{black}}
\else
\date{\today}
\fi
\fi
\begin{abstract}
The charged-particle pseudorapidity density measured over $4$ units of pseudorapidity in non-single-diffractive~(NSD) 
\pPb\ collisions at a centre-of-mass energy per nucleon pair \mbox{\snn = 5.02 TeV} is presented.
The average value at midrapidity is measured to be \final{\resetac} $\pm$ \final{\resetace} \syst,
which corresponds to \final{\resetacp} $\pm$ \final{\resetacpe}~\syst\ per participating nucleon,
calculated with the Glauber model.
This is \final{\facpp}\% lower than in NSD \pp\ collisions interpolated to the same collision energy,
and \final{\facdAu}\% higher than in \dAu\ collisions at $\snn=0.2$~TeV.
The measured pseudorapidity density in \pPb\ collisions is compared to model predictions, and provides 
new constraints on the description of particle production in high-energy nuclear collisions. 
\end{abstract}
\ifpreprint
\end{titlepage}
\setcounter{page}{2}
\else
\pacs{25.75.-q}
\maketitle
\fi
\ifdraft
\thispagestyle{fancyplain}
\fi

Particle production in proton--lead collisions, in contrast to \pp, is expected to be sensitive to 
nuclear effects in the initial state. 
In particular, coherence effects in the nuclear wave function are expected to influence the initial 
parton flux, as well as the underlying description of particle production in the scattering processes.
Therefore, measurements in \pPb\ collisions at the Large Hadron Collider (LHC) at CERN provide an essential 
experimental tool to discriminate between the initial and final state effects, and allow one to 
attribute the latter to the formation of hot QCD matter in heavy-ion collisions~\cite{Salgado:2011wc}.
Moreover, at LHC energies, the nuclear wave function is probed at small parton fractional momentum~$x$.
The growth of the parton densities with decreasing $x$ must be limited to satisfy unitarity bounds. 
One of the mechanisms providing such a limitation 
is often referred to as gluon saturation.
Its theoretical description varies between models of particle production resulting in significant 
differences in the predictions of the charged-particle pseudorapidity density.
Thus, the measurements of particle production in \pPb\ collisions constrain and potentially exclude 
certain models, and enhance the understanding of QCD at small $x$ and the initial state.

In this letter, the measurement of the primary charged-particle pseudorapidity density in \pPb\ collisions 
at a nucleon--nucleon centre-of-mass energy $\snn=5.02$ TeV with the \mbox{ALICE} detector~\cite{aliceapp} 
is reported.
The primary charged-particle density, $\dNdetal$, is measured in non single-diffractive~(NSD) \pPb\ 
collisions for $|\etal|<2$, where $\etal = - \ln \tan (\theta/2)$ and $\theta$ 
is the polar angle between the charged-particle direction and the beam axis~($z$).
Primary particles are defined as prompt particles produced in the collision, including decay products, 
except those from weak decays of strange particles. 
The data are compared to model predictions~\mbox{\cite{Dumitru:2011wq, Barnafoldi:2011px, 
Tribedy:2011aa, Xu:2012au, Albacete:2012xq}}, and to measurements in proton--nucleus~\mbox{\cite{Alber:1997sn,
Back:2003hx}}, NSD~\cite{ua1, ua51, ua52, starpp, cdf, alicepp2, Khachatryan:2010xs},
and inelastic~\cite{Thome:1977ky,Alner:1986xu,Aamodt:2010ft,Alver:2010ck} \pp~(\ppbar), as well as 
central heavy-ion~\cite{Abreu:2002fw, Adler:2001yq, Bearden:2001xw, Bearden:2001qq, Adcox:2000sp,
Back:2000gw, Back:2001bq, Back:2002wb, Alver:2010ck, Aamodt:2010pb, ATLAS:2011ag, Chatrchyan:2011pb} 
collisions.

The \pPb\ collisions were provided by the LHC during a short pilot run performed in September 2012
in preparation for the \pPb\ physics run scheduled for the beginning of 2013. 
The two-in-one magnet design of the LHC imposes the same magnetic rigidity of the beams in the two rings. 
Beam 1 consisted of protons at 4 TeV energy circulating in the negative $z$-direction in the ALICE laboratory
system, while beam 2 consisted of fully stripped $^{208}_{82}$Pb ions at $82\times4$ 
TeV energy circulating in the positive $z$-direction.
This configuration resulted in collisions at $\snn=5.02$~TeV in the nucleon--nucleon centre-of-mass system, 
which moves with a rapidity of $\Delta y_{\rm NN}=0.465$ in the direction of the proton beam.

The main detector for the present analysis is the Silicon Pixel Detector (\SPD), located in the inner barrel 
of the ALICE detector inside a solenoidal magnet providing a magnetic field of $0.5$~T.
The \SPD\ consists of two cylindrical layers of hybrid silicon pixel assemblies covering $|\etal|<2.0$ for the 
inner layer and $|\etal|<1.4$ for the outer layer with respect to vertices at the nominal interaction point.
A total of $9.8\times10^{6}$ pixels of size $50\times425$~$\mu$m$^2$ are read out, of which \final{93.5}\% were 
active during the run.
The primary trigger signal was provided by the \VZERO\ counters, two arrays of 32 scintillator 
tiles each covering the full azimuth within $2.8<\etal<5.1$~(\VZEROA) and $-3.7<\etal<-1.7$~(\VZEROC). 
The signal amplitude and arrival time collected in each scintillator are recorded.
The time resolution is better than \final{$1$}~ns, allowing discrimination of beam--beam collisions 
from background events produced outside of the interaction region.
Additionally, two neutron Zero Degree Calorimeters~(\ZDCs) are used, which are located at
\final{$+112.5$}~m~(\ZNA) and \final{$-112.5$}~m~(\ZNC) from the interaction point.
Their energy resolution is about \final{$20$}\% for single neutrons with a few TeV energy.
Each \ZDC\ also provided a trigger with high efficiency for single neutrons, 
which was used to collect a control sample of events for the estimation of the efficiency of the \VZERO\ trigger.

During the run, beams consisting of \final{13} bunches were circulating, with about 
\final{$10^{10}$} protons and \final{$6\times10^{7}$} Pb ions per bunch.
In the ALICE interaction region, \final{8} pairs of bunches were colliding, leading to a luminosity 
of \final{$8\times10^{25}$}~\lum.
The luminous region had a r.m.s.\ width of \final{$6.3$}~cm in the $z$-direction 
and about \final{60}~$\mu$m in the transverse direction.
The trigger was configured for high efficiency for hadronic events, requiring a signal in either 
\VZEROA\ or \VZEROC. 
This configuration led to an observed trigger rate of about \final{200}~Hz with a hadronic collision 
rate of about \final{150}~Hz.
In the offline analysis, a signal is required in both \VZEROA\ and \VZEROC. 
Beam--gas and other machine-induced background triggers with deposited energy above the thresholds in 
the \VZERO\ or \ZDC\ detectors are suppressed by requiring the arrival time to be compatible with that
of a nominal \pPb\ interaction.
The contamination from background is estimated from control triggers on 
non-colliding bunches, and found to be negligible.

In principle, the event sample obtained after these requirements consists of NSD collisions 
as well as single-diffractive~(SD) and electromagnetic~(EM) interactions.
The efficiency of the trigger and event selection on the different processes is estimated using a 
combination~(cocktail) of the following Monte Carlo~(MC) event generators:
a) DPMJET~\cite{Roesler:2000he} for NSD \pPb\ interactions,
b)~PHOJET~\cite{Engel:1995sb} tuned to \pp\ data at $\snn=2.76$ and $7$~TeV~\cite{abelev:2012sj}
together with a Glauber model~\cite{Alver:2008aq} for the contribution from SD interactions, and
c)~STARLIGHT~\cite{Djuvsland:2010qs} used together with PYTHIA~\cite{Sjostrand:2006za} or 
PHOJET~\cite{Engel:1995sb} for the proton excitation in the electromagnetic field of the $^{208}_{82}$Pb nucleus.
The DPMJET~\cite{Roesler:2000he} generator, which is based on the Gribov-Glauber approach and treats soft and hard 
scattering processes in an unified way, includes incoherent SD collisions of the projectile 
proton with target nucleons that are concentrated mainly on the surface of the nucleus.
These are removed by requiring that at least one of the binary nucleon--nucleon interactions is NSD. 
The relative weight of the events in the cocktail is given by the cross sections of the corresponding 
processes, which are taken to be \final{$2.0$}~b~(\final{$0.1$}~b) for NSD~(SD) 
collisions~(estimated from the Glauber model), and \final{$0.1$}--\final{$0.2$}~b for EM
interactions (estimated from STARLIGHT calculations).
The detector response to the cocktail is simulated using a model of the ALICE detector  
and the GEANT3 simulation tool~\cite{geant3ref2}.
An efficiency of \final{$99.2$}\% for NSD collisions and a negligible contamination from
SD and EM interactions are obtained.

From the collected data sample used for the analysis, \final{$0.8\times10^6$} events pass the selection criteria.
Among the selected events, \final{$98.5$}\% are found to have a primary vertex. 
The corresponding fraction in DPMJET~\cite{Roesler:2000he} for NSD collisions is \final{$99.4$}\%
with the probability of selecting an event without a primary vertex of \final{$41$}\%. 
Taking into account the difference of the fraction of events without vertex in the data and 
the simulation results in an overall selection  efficiency of \final{$96.4$}\% for 
NSD events entering the analysis.

The $\dNdetal$ analysis techniques employed are identical to those described in \Ref{Aamodt:2010pb}, 
where the similar measurement is reported for \PbPb\ collisions. 
Events are selected with a reconstructed vertex within $|z_{\rm vtx}|<$~\final{18}~cm,
which results in a $|\etal|<2$ coverage for the $\dNdetal$ measurement. 
Tracklet candidates are formed using the position of the primary vertex and two hits, one on each \SPD\ layer.
From these candidates, tracklets are selected by a requirement on the sum of the squares of 
the differences~(residuals) in azimuthal and polar angles relative to the primary vertex for each hit,
effectively selecting charged particles with transverse momentum~(\pt) above 50~MeV/$c$, while
particles below 50~MeV/$c$ are mostly absorbed by detector material. 
The charged-particle pseudorapidity density is then obtained from the measured distribution of tracklets
$\dNdetatrl$ as $\dNdetal = \alpha \, (1-\beta)\,\dNdetatrl$.
The correction $\alpha$ accounts for the acceptance and efficiency for a primary particle to produce a tracklet, 
while $\beta$ is the contamination of reconstructed tracklets from combinations of hits not produced by the 
same primary particle. 
Both are determined as a function of the $z$-position of the primary vertex and the pseudorapidity of the tracklet
from detector simulations using DPMJET~\cite{Roesler:2000he} and GEANT3~\cite{geant3ref2},
and found to be on average \final{1.2} and \final{0.01}, respectively.
Since the corrections applied in the analysis implicitly only account for the fraction of events without
vertex given by the simulation, the $\dNdetal$ is further corrected by \final{$-2.2$}\% for the difference 
of this fraction in the data and the simulation.

The following sources of systematic uncertainties have been considered.
The uncertainty in detector acceptance is estimated to be \final{$1.5$}\% determined from the change of the 
multiplicity at a given $\etal$ by varying the range of the $z$-position of the vertex.
The uncertainties resulting from the subtraction of the combinatorial background and from the contribution of 
weak decays are estimated to be \final{$0.3$}\% and \final{$0.8$}\%, respectively. They are determined from the 
comparison in data and simulation of the tracklet residual distributions, in which the tails are dominated 
by combinatorial background and secondaries.
The uncertainty due to the particle composition is estimated to be \final{$1$}\%, which was determined by changing 
the relative abundances of pions, kaons and protons by a factor of \final{$2$} in the simulation.
The uncertainty due to the correction down to zero \pt\ is estimated to be \final{$1$}\% 
by varying the amount of undetected particles at low $\pt$ by \final{$50$}\%.
The uncertainty related to the trigger and event selection efficiency for NSD collisions is
estimated to be \final{$3.1$}\% using a small sample of events collected with the \ZNA\ trigger
with an offline selection on the deposited energy corresponding to approximately $12$ neutrons 
from the Pb remnant.
The value used for the threshold has been determined from DPMJET with associated nuclear fragment 
production~\cite{Andersen20041302}, and was chosen to suppress the contamination of the EM and SD interactions.
In total, a systematic uncertainty of about \final{$3.8$}\% is obtained by adding in quadrature all the 
contributions. 

\begin{figure}[tbh!f]
\begin{center}
\includegraphics[width=\linewidth]{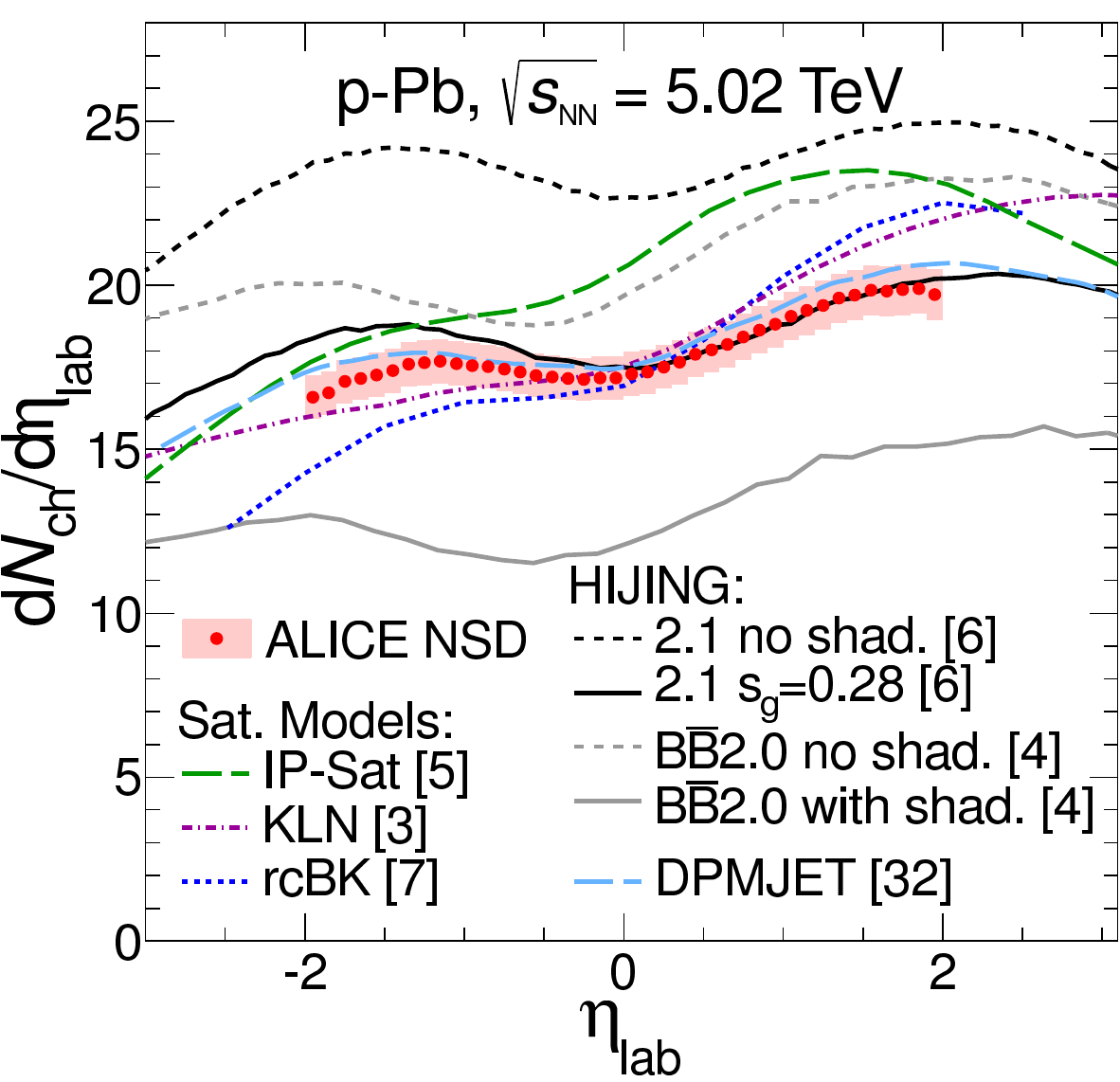}
\caption{\label{fig1} 
Pseudorapidity density of charged particles measured in NSD \pPb\ collisions at $\snn=5.02$~TeV
compared to theoretical predictions~\cite{Dumitru:2011wq, Barnafoldi:2011px, Tribedy:2011aa, Xu:2012au, 
Albacete:2012xq}. The calculations~\cite{Barnafoldi:2011px, Tribedy:2011aa} have been 
shifted to the laboratory system.}
\end{center}
\end{figure}

\begin{table}[tbh!f]
\begin{center}
\begin{tabular}{l*{4}{c}}
 & \multicolumn{3}{c}{$\dNdetal$} & \\
 & $-2.0$ & $0.0$ & $2.0$ & \multirow{-2}{3cm}{$\frac{\left. \dNdetal \right|_{\etal = 2.0}}{\left. \dNdetal \right|_{\etal = -2.0}}$} \\
\hline
ALICE & $16.65$ & $17.24$ & $19.81$ & $1.19$ \\
 & $\pm 0.65$ & $\pm 0.66$ & $\pm 0.78$ & $\pm 0.05$ \\
\hline
Saturation Models & & & & \\
IP-Sat~\cite{Tribedy:2011aa} & 17.55 & 20.55 & 23.11 & 1.32 \\
KLN~\cite{Dumitru:2011wq} & 15.96 & 17.51 & 22.02 & 1.38 \\
rcBK~\cite{Albacete:2012xq} & 14.27 & 16.94 & 22.51 & 1.58 \\
\hline
HIJING & & & & \\
2.1 no shad.~\cite{Xu:2012au} & 23.58 & 22.67 & 24.96 & 1.06 \\
2.1 $s_{g}=0.28$~\cite{Xu:2012au} & 18.30 & 17.49 & 20.21 & 1.10 \\
B$\bar{\mathrm{B}}$2.0 no shad.~\cite{Barnafoldi:2011px} & 20.03 & 19.68 & 23.24 & 1.16 \\
B$\bar{\mathrm{B}}$2.0 with shad.~\cite{Barnafoldi:2011px} & 12.97 & 12.09 & 15.16 & 1.17 \\
\hline
DPMJET~\cite{Roesler:2000he} & 17.50 & 17.61 & 20.67 & 1.18 \\
\end{tabular}
\caption{Comparison of the pseudorapidity distribution between data and the models at $\etal=-2$, $0$ and $2$
(integrated in $0.2$ units of pseudorapidity) as well as the ratio of $\dNdetal$ at $\etal=2$ to that at $\etal=-2$.
The uncertainty introduced by taking the ratio neglecting the Jacobian amounts to about $2$ and $6$\% 
estimated for the saturation and HIJING models, respectively.}
\label{table1}
\end{center}
\end{table}

The resulting pseudorapidity density is presented in \Fig{fig1} for $|\etal|<2$. 
A forward--backward asymmetry between the proton and lead hemispheres is clearly visible.
The measurement is compared to particle production models~\cite{Dumitru:2011wq, Barnafoldi:2011px, Tribedy:2011aa, 
Xu:2012au, Albacete:2012xq} that describe similar measurements in other collision systems~\cite{Back:2003hx, 
Abreu:2002fw, Adler:2001yq, Bearden:2001xw, Bearden:2001qq, Adcox:2000sp, Back:2000gw, Back:2001bq, Back:2002wb, 
Alver:2010ck, Aamodt:2010pb, ATLAS:2011ag, Chatrchyan:2011pb}. 
The two-component models~\cite{Barnafoldi:2011px,Xu:2012au} combine perturbative QCD processes
with soft interactions, and include nuclear modification of the initial parton distributions.
The saturation models~~\cite{Dumitru:2011wq, Tribedy:2011aa, Albacete:2012xq}
employ coherence effects to reduce the number of soft gluons available for particle production 
below a given energy scale.
The calculations~\cite{Dumitru:2011wq, Xu:2012au, Albacete:2012xq} at $\snn=5.02$~TeV were provided 
by the authors in the laboratory system. 
The calculations that were performed in the centre-of-mass system~\cite{Barnafoldi:2011px, 
Tribedy:2011aa} have been shifted by $\Delta y_{\rm NN}$ into the ALICE laboratory system.
For low-\pt\ particles, this is only an approximation of a Lorentz transformation. 
In the $\etal$-range of our measurement, the error on the $\dNdetal$ density induced by this procedure 
is estimated using the HIJING model~\cite{hijing}, and found to be below \final{6}\%.
It is worth noting that the HIJING calculations include single-diffraction, which from the HIJING 
generator~\cite{hijing} is estimated to be about \final{$4$}\%.
A comparison of the model calculations with the data shows that most of the models that include 
shadowing~\cite{Xu:2012au} or saturation~\cite{Dumitru:2011wq, Albacete:2012xq} predict the measured 
multiplicity values to within \final{20}\%~(see also Tab.~\ref{table1}).
The HIJING/B$\overline{\rm B}$$2.0$~\cite{Barnafoldi:2011px} model, which uses an energy and nuclear thickness 
dependent string tension to mimic the effect of strong longitudinal color fields, predicts values below the data 
when including shadowing, and above the data when excluding shadowing.
DPMJET~\cite{Roesler:2000he} (normalized to NSD) and HIJING 2.1~\cite{Xu:2012au}, where the gluon shadowing 
parameter $s_g=0.28$ was tuned to describe experimental data on rapidity distributions in \dAu\ collisions at 
$\snn=0.2$ TeV~(RHIC)~\cite{Back:2003hx,Alver:2010ck}, give values that are closest to the data.
Both also describe the pseudorapidity shape relatively well, whereas the saturation models~\cite{Dumitru:2011wq, 
Tribedy:2011aa, Albacete:2012xq} exhibit a steeper $\etal$ dependence than the data.
This can also be seen in Tab.~\ref{table1} by quantifying the density at midrapidity, near the proton and lead peak 
regions, as well as the ratio of $\dNdetal$ at $\etal=2$ to that at $\etal=-2$, for the data~(integrated in $0.2$
units of pseudorapidity) and the models. The error introduced by taking the ratio amounts to about $2$ and $6$\% 
for the saturation and HIJING models.

The charged-particle pseudorapidity density at midrapidity in the laboratory system~($\abs{\etal}<0.5$)
is $\dNdetal=$ \final{\resetal} $\pm$ \final{0.01} \stat\ $\pm$ \final{\resetale} \syst.
The statistical uncertainty is neglected in the following.
To obtain the pseudorapidity density in the centre-of-mass system, the data is integrated
in the range $-0.965 < \etal < 0.035$, and corrected for the effect of the $\Delta y$ shift.
The correction is estimated from the HIJING model~\cite{hijing} to be \final{3}\%, with an 
uncertainty of \final{1.5}\%, added in quadrature to the systematic uncertainty.
The resulting pseudorapidity density in the nucleon--nucleon centre-of-mass system is 
$\dNdetac=$ \final{\resetac} $\pm$ \final{\resetace} \syst.

\begin{figure}[tbh!]
\begin{center}
\includegraphics[width=\linewidth]{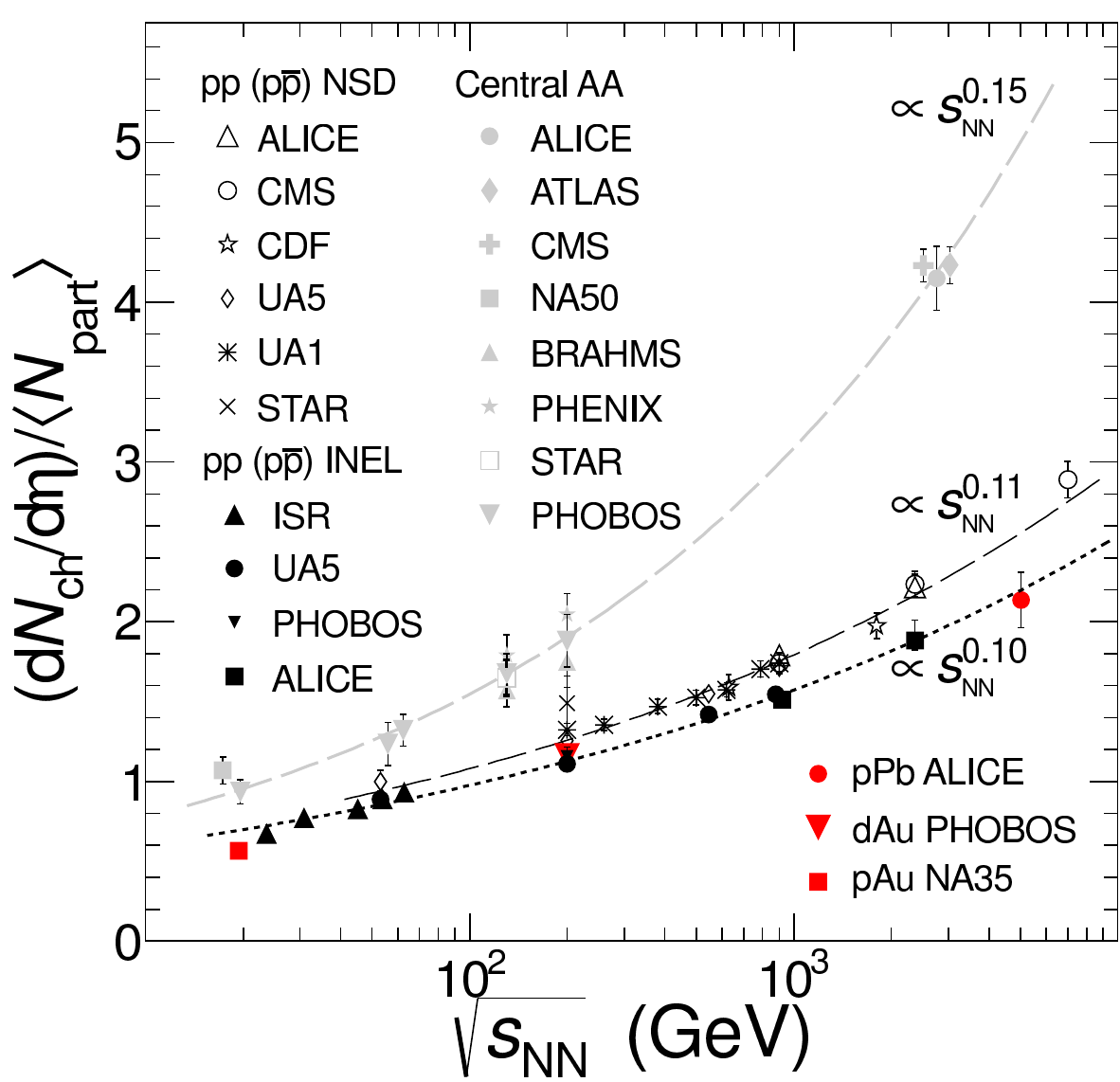}
\caption{\label{fig2}
Charged-particle pseudorapidity density at midrapidity normalized to the number of participants,
calculated with the Glauber model,
for \pPb, \pAu\ and \dAu~\cite{Alber:1997sn,Back:2003hx} collisions as a function of $\snn$, compared
to NSD~\cite{ua1, ua51, ua52, starpp, cdf, alicepp2, Khachatryan:2010xs},
and inelastic~\cite{Thome:1977ky,Alner:1986xu,Aamodt:2010ft,Alver:2010ck} \pp~(\ppbar) collisions, 
as well as central heavy-ion~\cite{Abreu:2002fw, Adler:2001yq, Bearden:2001xw, Bearden:2001qq, Adcox:2000sp,
Back:2000gw, Back:2001bq, Back:2002wb, Alver:2010ck, Aamodt:2010pb, ATLAS:2011ag, Chatrchyan:2011pb} 
collisions.
The curves $\propto s_{\rm NN}^{0.11}$ and $s_{\rm NN}^{0.15}$~(from \cite{Aamodt:2010pb}) are superimposed 
on the NSD \pp~(\ppbar) and central heavy-ion data, respectively, while 
$\propto s_{\rm NN}^{0.10}$~(from \cite{Aamodt:2010ft}) on the inelastic \pp~(\ppbar) data.}
\end{center}
\end{figure}

In order to compare bulk particle production in different collision systems, the charged particle density is scaled 
by the number of participating nucleons, determined using the Glauber model~\cite{Alver:2008aq}
with a nuclear radius of $6.62\pm 0.06$~fm and a skin depth of $0.546 \pm 0.010$~fm, a hard-sphere exclusion distance 
of $0.4 \pm 0.4$~fm for the lead nucleus, a radius of $0.6 \pm 0.2$~fm for the proton, and an inelastic 
nucleon--nucleon cross section of $70 \pm 5$~mb. 
The latter is obtained by interpolating data at different centre-of-mass energies~\cite{Nakamura:2010zzi} 
including measurements at 2.76 and 7 TeV~\cite{abelev:2012sj,Antchev:2011vs}.
The number of participants for minimum-bias events is found to be distributed with an average
$\avNpart = $ \final{7.9} $\pm$ \final{0.6} and an r.m.s.\ width of \final{5.1}. 
The uncertainty of \final{7.6}\% on $\avNpart$ is obtained by varying the parameters of the Glauber calculation 
within the ranges mentioned above~(as explained in \Ref{Aamodt:2010cz}).
Note that the number of participants would increase by only \final{$2.5$}\% if normalized to NSD events in the 
Glauber calculation. 
Normalizing to the number of participants gives $(\dNdetac)/\avNpart= $ \final{\resetacp} $\pm$ 
\final{\resetacpe}~\syst. 
In \Fig{fig2}, this value is compared to measurements in \pAu\ and \dAu~\cite{Alber:1997sn, Back:2003hx} 
collisions, NSD~\cite{ua1, ua51, ua52, starpp, cdf, alicepp2, Khachatryan:2010xs},
and inelastic~\cite{Thome:1977ky, Alner:1986xu, Aamodt:2010ft, Alver:2010ck} \pp~(\ppbar), as well as 
central heavy-ion~\cite{Abreu:2002fw, Adler:2001yq, Bearden:2001xw, Bearden:2001qq, Adcox:2000sp,
Back:2000gw, Back:2001bq, Back:2002wb, Alver:2010ck, Aamodt:2010pb, ATLAS:2011ag, Chatrchyan:2011pb} 
collisions, over a wide range of collision energies. (Data for \dAu\ at $\snn=200$ GeV from \cite{Arsene:2004cn,Abelev:2007nt} 
are consistent with that from \cite{Back:2003hx} and not shown in the figure.)
The $(\dNdetac)/\avNpart$ at $\snn=5.02$~TeV is found to be \final{\facpp}\% lower than in NSD \pp\
and consistent with inelastic \pp\ collisions interpolated to $\snn=5.02$~TeV, and \final{\facdAu}\% 
higher than in \dAu\ collisions at $\snn=0.2$~TeV.

In summary, the charged-particle pseudorapidity density in $|\etal|<2$ in non-single-diffractive \pPb\ 
collisions at $\snn = 5.02$~TeV is presented.
At midrapidity, $\dNdetac = $ \final{\resetac} $\pm$ \final{\resetace}~\syst\ is measured, corresponding to 
\final{\resetacp} $\pm$ \final{\resetacpe}~\syst\ charged particles per unit pseudorapidity per 
participant, where the number of participants are calculated with the Glauber model.
The new measurement extends the study of charged-particle densities in proton--nucleus collisions 
into the TeV scale, and provides new constraints on the description of particle production in 
high-energy nuclear collisions. 

\ifpreprint
\iffull
\newenvironment{acknowledgement}{\relax}{\relax}
\begin{acknowledgement}
\section*{Acknowledgements}
The ALICE collaboration would like to thank J.\ Albacete, A.\ Dumitru, M.\ Gyulassy, A.\ Rezaeian, 
V.\ Topor,  P.\ Tribedy and X.-N.\ Wang for helpful discussions and their model predictions.
\\
The ALICE collaboration would like to thank all its engineers and technicians for their invaluable contributions to the construction of the experiment and the CERN accelerator teams for the outstanding performance of the LHC complex.
\\
The ALICE collaboration acknowledges the following funding agencies for their support in building and
running the ALICE detector:
 \\
State Committee of Science, Calouste Gulbenkian Foundation from
Lisbon and Swiss Fonds Kidagan, Armenia;
 \\
Conselho Nacional de Desenvolvimento Cient\'{\i}fico e Tecnol\'{o}gico (CNPq), Financiadora de Estudos e Projetos (FINEP),
Funda\c{c}\~{a}o de Amparo \`{a} Pesquisa do Estado de S\~{a}o Paulo (FAPESP);
 \\
National Natural Science Foundation of China (NSFC), the Chinese Ministry of Education (CMOE)
and the Ministry of Science and Technology of China (MSTC);
 \\
Ministry of Education and Youth of the Czech Republic;
 \\
Danish Natural Science Research Council, the Carlsberg Foundation and the Danish National Research Foundation;
 \\
The European Research Council under the European Community's Seventh Framework Programme;
 \\
Helsinki Institute of Physics and the Academy of Finland;
 \\
French CNRS-IN2P3, the `Region Pays de Loire', `Region Alsace', `Region Auvergne' and CEA, France;
 \\
German BMBF and the Helmholtz Association;
\\
General Secretariat for Research and Technology, Ministry of
Development, Greece;
\\
Hungarian OTKA and National Office for Research and Technology (NKTH);
 \\
Department of Atomic Energy and Department of Science and Technology of the Government of India;
 \\
Istituto Nazionale di Fisica Nucleare (INFN) of Italy;
 \\
MEXT Grant-in-Aid for Specially Promoted Research, Ja\-pan;
 \\
Joint Institute for Nuclear Research, Dubna;
 \\
National Research Foundation of Korea (NRF);
 \\
CONACYT, DGAPA, M\'{e}xico, ALFA-EC and the HELEN Program (High-Energy physics Latin-American--European Network);
 \\
Stichting voor Fundamenteel Onderzoek der Materie (FOM) and the Nederlandse Organisatie voor Wetenschappelijk Onderzoek (NWO), Netherlands;
 \\
Research Council of Norway (NFR);
 \\
Polish Ministry of Science and Higher Education;
 \\
National Authority for Scientific Research - NASR (Autoritatea Na\c{t}ional\u{a} pentru Cercetare \c{S}tiin\c{t}ific\u{a} - ANCS);
 \\
Ministry of Education and Science of Russian Federation,
International Science and Technology Center, Russian Academy of
Sciences, Russian Federal Agency of Atomic Energy, Russian Federal
Agency for Science and Innovations and CERN-INTAS;
 \\
Ministry of Education of Slovakia;
 \\
Department of Science and Technology, South Africa;
 \\
CIEMAT, EELA, Ministerio de Educaci\'{o}n y Ciencia of Spain, Xunta de Galicia (Conseller\'{\i}a de Educaci\'{o}n),
CEA\-DEN, Cubaenerg\'{\i}a, Cuba, and IAEA (International Atomic Energy Agency);
 \\
Swedish Research Council (VR) and Knut $\&$ Alice Wallenberg
Foundation (KAW);
 \\
Ukraine Ministry of Education and Science;
 \\
United Kingdom Science and Technology Facilities Council (STFC);
 \\
The United States Department of Energy, the United States National
Science Foundation, the State of Texas, and the State of Ohio.
\end{acknowledgement}
\fi
\ifbibtex
\bibliographystyle{utphys}
\bibliography{multpap}{}
\else
\providecommand{\href}[2]{#2}\begingroup\raggedright\endgroup

\fi
\iffull
\newpage
\appendix
\section{The ALICE Collaboration}
\label{app:collab}

\begingroup
\small
\begin{flushleft}
B.~Abelev\Irefn{org1234}\And
J.~Adam\Irefn{org1274}\And
D.~Adamov\'{a}\Irefn{org1283}\And
A.M.~Adare\Irefn{org1260}\And
M.M.~Aggarwal\Irefn{org1157}\And
G.~Aglieri~Rinella\Irefn{org1192}\And
M.~Agnello\Irefn{org1313}\And
A.G.~Agocs\Irefn{org1143}\And
A.~Agostinelli\Irefn{org1132}\And
Z.~Ahammed\Irefn{org1225}
% ...
\renewcommand\labelenumi{\textsuperscript{\theenumi}~}
\section*{Affiliation notes}
\renewcommand\theenumi{\roman{enumi}}
\begin{Authlist}
\item \Adef{0}Deceased
\item \Adef{M.V.Lomonosov}Also at: M.V.Lomonosov Moscow State University, D.V.Skobeltsyn Institute of Nuclear Physics, Moscow, Russia
\item \Adef{University of Belgrade, Faculty of Physics and "Vinvca" Institute of Nuclear Sciences, Belgrade, Serbia}Also at: University of Belgrade, Faculty of Physics and "Vinvca" Institute of Nuclear Sciences, Belgrade, Serbia
\end{Authlist}
\section*{Collaboration Institutes}
\renewcommand\theenumi{\arabic{enumi}~}
\begin{Authlist}
\item \Idef{org1332}A. I. Alikhanyan National Science Laboratory (Yerevan Physics Institute) Foundation, Yerevan, Armenia
\item \Idef{org1279}Benem\'{e}rita Universidad Aut\'{o}noma de Puebla, Puebla, Mexico
\item \Idef{org1220}Bogolyubov Institute for Theoretical Physics, Kiev, Ukraine
\item \Idef{org20959}Bose Institute, Department of Physics and Centre for Astroparticle Physics and Space Science (CAPSS), Kolkata, India
\item \Idef{org1262}Budker Institute for Nuclear Physics, Novosibirsk, Russia
\item \Idef{org1292}California Polytechnic State University, San Luis Obispo, California, United States
\item \Idef{org1329}Central China Normal University, Wuhan, China
\item \Idef{org14939}Centre de Calcul de l'IN2P3, Villeurbanne, France
\item \Idef{org1197}Centro de Aplicaciones Tecnol\'{o}gicas y Desarrollo Nuclear (CEADEN), Havana, Cuba
\item \Idef{org1242}Centro de Investigaciones Energ\'{e}ticas Medioambientales y Tecnol\'{o}gicas (CIEMAT), Madrid, Spain
\item \Idef{org1244}Centro de Investigaci\'{o}n y de Estudios Avanzados (CINVESTAV), Mexico City and M\'{e}rida, Mexico
\item \Idef{org1335}Centro Fermi -- Centro Studi e Ricerche e Museo Storico della Fisica ``Enrico Fermi'', Rome, Italy
\item \Idef{org17347}Chicago State University, Chicago, United States
\item \Idef{org1288}Commissariat \`{a} l'Energie Atomique, IRFU, Saclay, France
\item \Idef{org15782}COMSATS Institute of Information Technology (CIIT), Islamabad, Pakistan
\item \Idef{org1294}Departamento de F\'{\i}sica de Part\'{\i}culas and IGFAE, Universidad de Santiago de Compostela, Santiago de Compostela, Spain
\item \Idef{org1106}Department of Physics Aligarh Muslim University, Aligarh, India
\item \Idef{org1121}Department of Physics and Technology, University of Bergen, Bergen, Norway
\item \Idef{org1162}Department of Physics, Ohio State University, Columbus, Ohio, United States
\item \Idef{org1300}Department of Physics, Sejong University, Seoul, South Korea
\item \Idef{org1268}Department of Physics, University of Oslo, Oslo, Norway
\item \Idef{org1312}Dipartimento di Fisica dell'Universit\`{a} and Sezione INFN, Turin, Italy
\item \Idef{org1132}Dipartimento di Fisica dell'Universit\`{a} and Sezione INFN, Bologna, Italy
\item \Idef{org1145}Dipartimento di Fisica dell'Universit\`{a} and Sezione INFN, Cagliari, Italy
\item \Idef{org1315}Dipartimento di Fisica dell'Universit\`{a} and Sezione INFN, Trieste, Italy
\item \Idef{org1285}Dipartimento di Fisica dell'Universit\`{a} `La Sapienza' and Sezione INFN, Rome, Italy
\item \Idef{org1154}Dipartimento di Fisica e Astronomia dell'Universit\`{a} and Sezione INFN, Catania, Italy
\item \Idef{org1270}Dipartimento di Fisica e Astronomia dell'Universit\`{a} and Sezione INFN, Padova, Italy
\item \Idef{org1290}Dipartimento di Fisica `E.R.~Caianiello' dell'Universit\`{a} and Gruppo Collegato INFN, Salerno, Italy
\item \Idef{org1103}Dipartimento di Scienze e Innovazione Tecnologica dell'Universit\`{a} del Piemonte Orientale and Gruppo Collegato INFN, Alessandria, Italy
\item \Idef{org1114}Dipartimento Interateneo di Fisica `M.~Merlin' and Sezione INFN, Bari, Italy
\item \Idef{org1237}Division of Experimental High Energy Physics, University of Lund, Lund, Sweden
\item \Idef{org1192}European Organization for Nuclear Research (CERN), Geneva, Switzerland
\item \Idef{org1227}Fachhochschule K\"{o}ln, K\"{o}ln, Germany
\item \Idef{org1122}Faculty of Engineering, Bergen University College, Bergen, Norway
\item \Idef{org1136}Faculty of Mathematics, Physics and Informatics, Comenius University, Bratislava, Slovakia
\item \Idef{org1274}Faculty of Nuclear Sciences and Physical Engineering, Czech Technical University in Prague, Prague, Czech Republic
\item \Idef{org1229}Faculty of Science, P.J.~\v{S}af\'{a}rik University, Ko\v{s}ice, Slovakia
\item \Idef{org1184}Frankfurt Institute for Advanced Studies, Johann Wolfgang Goethe-Universit\"{a}t Frankfurt, Frankfurt, Germany
\item \Idef{org1215}Gangneung-Wonju National University, Gangneung, South Korea
\item \Idef{org20958}Gauhati University, Department of Physics, Guwahati, India
\item \Idef{org1212}Helsinki Institute of Physics (HIP) and University of Jyv\"{a}skyl\"{a}, Jyv\"{a}skyl\"{a}, Finland
\item \Idef{org1203}Hiroshima University, Hiroshima, Japan
\item \Idef{org1254}Indian Institute of Technology Bombay (IIT), Mumbai, India
\item \Idef{org36378}Indian Institute of Technology Indore, Indore, India (IITI)
\item \Idef{org1266}Institut de Physique Nucl\'{e}aire d'Orsay (IPNO), Universit\'{e} Paris-Sud, CNRS-IN2P3, Orsay, France
\item \Idef{org1277}Institute for High Energy Physics, Protvino, Russia
\item \Idef{org1249}Institute for Nuclear Research, Academy of Sciences, Moscow, Russia
\item \Idef{org1320}Nikhef, National Institute for Subatomic Physics and Institute for Subatomic Physics of Utrecht University, Utrecht, Netherlands
\item \Idef{org1250}Institute for Theoretical and Experimental Physics, Moscow, Russia
\item \Idef{org1230}Institute of Experimental Physics, Slovak Academy of Sciences, Ko\v{s}ice, Slovakia
\item \Idef{org1127}Institute of Physics, Bhubaneswar, India
\item \Idef{org1275}Institute of Physics, Academy of Sciences of the Czech Republic, Prague, Czech Republic
\item \Idef{org1139}Institute of Space Sciences (ISS), Bucharest, Romania
\item \Idef{org27399}Institut f\"{u}r Informatik, Johann Wolfgang Goethe-Universit\"{a}t Frankfurt, Frankfurt, Germany
\item \Idef{org1185}Institut f\"{u}r Kernphysik, Johann Wolfgang Goethe-Universit\"{a}t Frankfurt, Frankfurt, Germany
\item \Idef{org1177}Institut f\"{u}r Kernphysik, Technische Universit\"{a}t Darmstadt, Darmstadt, Germany
\item \Idef{org1256}Institut f\"{u}r Kernphysik, Westf\"{a}lische Wilhelms-Universit\"{a}t M\"{u}nster, M\"{u}nster, Germany
\item \Idef{org1246}Instituto de Ciencias Nucleares, Universidad Nacional Aut\'{o}noma de M\'{e}xico, Mexico City, Mexico
\item \Idef{org1247}Instituto de F\'{\i}sica, Universidad Nacional Aut\'{o}noma de M\'{e}xico, Mexico City, Mexico
\item \Idef{org23333}Institut of Theoretical Physics, University of Wroclaw
\item \Idef{org1308}Institut Pluridisciplinaire Hubert Curien (IPHC), Universit\'{e} de Strasbourg, CNRS-IN2P3, Strasbourg, France
\item \Idef{org1182}Joint Institute for Nuclear Research (JINR), Dubna, Russia
\item \Idef{org1143}KFKI Research Institute for Particle and Nuclear Physics, Hungarian Academy of Sciences, Budapest, Hungary
\item \Idef{org1199}Kirchhoff-Institut f\"{u}r Physik, Ruprecht-Karls-Universit\"{a}t Heidelberg, Heidelberg, Germany
\item \Idef{org20954}Korea Institute of Science and Technology Information, Daejeon, South Korea
\item \Idef{org1160}Laboratoire de Physique Corpusculaire (LPC), Clermont Universit\'{e}, Universit\'{e} Blaise Pascal, CNRS--IN2P3, Clermont-Ferrand, France
\item \Idef{org1194}Laboratoire de Physique Subatomique et de Cosmologie (LPSC), Universit\'{e} Joseph Fourier, CNRS-IN2P3, Institut Polytechnique de Grenoble, Grenoble, France
\item \Idef{org1187}Laboratori Nazionali di Frascati, INFN, Frascati, Italy
\item \Idef{org1232}Laboratori Nazionali di Legnaro, INFN, Legnaro, Italy
\item \Idef{org1125}Lawrence Berkeley National Laboratory, Berkeley, California, United States
\item \Idef{org1234}Lawrence Livermore National Laboratory, Livermore, California, United States
\item \Idef{org1251}Moscow Engineering Physics Institute, Moscow, Russia
\item \Idef{org1322}National Centre for Nuclear Studies, Warsaw, Poland
\item \Idef{org1140}National Institute for Physics and Nuclear Engineering, Bucharest, Romania
\item \Idef{org1165}Niels Bohr Institute, University of Copenhagen, Copenhagen, Denmark
\item \Idef{org1109}Nikhef, National Institute for Subatomic Physics, Amsterdam, Netherlands
\item \Idef{org1283}Nuclear Physics Institute, Academy of Sciences of the Czech Republic, \v{R}e\v{z} u Prahy, Czech Republic
\item \Idef{org1264}Oak Ridge National Laboratory, Oak Ridge, Tennessee, United States
\item \Idef{org1189}Petersburg Nuclear Physics Institute, Gatchina, Russia
\item \Idef{org1170}Physics Department, Creighton University, Omaha, Nebraska, United States
\item \Idef{org1157}Physics Department, Panjab University, Chandigarh, India
\item \Idef{org1112}Physics Department, University of Athens, Athens, Greece
\item \Idef{org1152}Physics Department, University of Cape Town and  iThemba LABS, National Research Foundation, Somerset West, South Africa
\item \Idef{org1209}Physics Department, University of Jammu, Jammu, India
\item \Idef{org1207}Physics Department, University of Rajasthan, Jaipur, India
\item \Idef{org1200}Physikalisches Institut, Ruprecht-Karls-Universit\"{a}t Heidelberg, Heidelberg, Germany
\item \Idef{org1325}Purdue University, West Lafayette, Indiana, United States
\item \Idef{org1281}Pusan National University, Pusan, South Korea
\item \Idef{org1176}Research Division and ExtreMe Matter Institute EMMI, GSI Helmholtzzentrum f\"ur Schwerionenforschung, Darmstadt, Germany
\item \Idef{org1334}Rudjer Bo\v{s}kovi\'{c} Institute, Zagreb, Croatia
\item \Idef{org1298}Russian Federal Nuclear Center (VNIIEF), Sarov, Russia
\item \Idef{org1252}Russian Research Centre Kurchatov Institute, Moscow, Russia
\item \Idef{org1224}Saha Institute of Nuclear Physics, Kolkata, India
\item \Idef{org1130}School of Physics and Astronomy, University of Birmingham, Birmingham, United Kingdom
\item \Idef{org1338}Secci\'{o}n F\'{\i}sica, Departamento de Ciencias, Pontificia Universidad Cat\'{o}lica del Per\'{u}, Lima, Peru
\item \Idef{org1133}Sezione INFN, Bologna, Italy
\item \Idef{org1155}Sezione INFN, Catania, Italy
\item \Idef{org1271}Sezione INFN, Padova, Italy
\item \Idef{org1146}Sezione INFN, Cagliari, Italy
\item \Idef{org1313}Sezione INFN, Turin, Italy
\item \Idef{org1316}Sezione INFN, Trieste, Italy
\item \Idef{org1286}Sezione INFN, Rome, Italy
\item \Idef{org1115}Sezione INFN, Bari, Italy
\item \Idef{org36377}Nuclear Physics Group, STFC Daresbury Laboratory, Daresbury, United Kingdom
\item \Idef{org1258}SUBATECH, Ecole des Mines de Nantes, Universit\'{e} de Nantes, CNRS-IN2P3, Nantes, France
\item \Idef{org1304}Technical University of Split FESB, Split, Croatia
\item \Idef{org1168}The Henryk Niewodniczanski Institute of Nuclear Physics, Polish Academy of Sciences, Cracow, Poland
\item \Idef{org17361}The University of Texas at Austin, Physics Department, Austin, TX, United States
\item \Idef{org1173}Universidad Aut\'{o}noma de Sinaloa, Culiac\'{a}n, Mexico
\item \Idef{org1296}Universidade de S\~{a}o Paulo (USP), S\~{a}o Paulo, Brazil
\item \Idef{org1149}Universidade Estadual de Campinas (UNICAMP), Campinas, Brazil
\item \Idef{org1239}Universit\'{e} de Lyon, Universit\'{e} Lyon 1, CNRS/IN2P3, IPN-Lyon, Villeurbanne, France
\item \Idef{org1205}University of Houston, Houston, Texas, United States
\item \Idef{org20371}University of Technology and Austrian Academy of Sciences, Vienna, Austria
\item \Idef{org1222}University of Tennessee, Knoxville, Tennessee, United States
\item \Idef{org1310}University of Tokyo, Tokyo, Japan
\item \Idef{org1318}University of Tsukuba, Tsukuba, Japan
\item \Idef{org21360}Eberhard Karls Universit\"{a}t T\"{u}bingen, T\"{u}bingen, Germany
\item \Idef{org1225}Variable Energy Cyclotron Centre, Kolkata, India
\item \Idef{org1306}V.~Fock Institute for Physics, St. Petersburg State University, St. Petersburg, Russia
\item \Idef{org1323}Warsaw University of Technology, Warsaw, Poland
\item \Idef{org1179}Wayne State University, Detroit, Michigan, United States
\item \Idef{org1260}Yale University, New Haven, Connecticut, United States
\item \Idef{org15649}Yildiz Technical University, Istanbul, Turkey
\item \Idef{org1301}Yonsei University, Seoul, South Korea
\item \Idef{org1327}Zentrum f\"{u}r Technologietransfer und Telekommunikation (ZTT), Fachhochschule Worms, Worms, Germany
\end{Authlist}
\endgroup

\fi
\else 
\iffull
\\

\input{references_prl.tex}
\else
\ifbibtex
\bibliographystyle{apsrev4-1}
\bibliography{multpap}{}
\else
\input{references_prl.tex}
\fi
\fi
\fi
\end{document}